# Strain and Vector-Magnetic-Field Tuning of the Anomalous Phase in $Sr_3Ru_2O_7$.

Daniel O. Brodsky[1,2], Mark E. Barber[1,2], Jan A. N. Bruin[2,3], Rodolfo A. Borzi[4], Santiago A. Grigera[2,4], Robin S. Perry[5], Andrew P. Mackenzie[1,2,*], Clifford W. Hicks[1,§]

December 2016

[1]Max Planck Institute for Chemical Physics of Solids, Nöthnitzer Straße 40, 01187 Dresden, Germany
[2]Scottish Universities Physics Alliance (SUPA), School of Physics and Astronomy, North Haugh,
University of St. Andrews, St. Andrews KY16 9SS, United Kingdom
[3]Max Planck Institute for Solid State Physics, Heisenbergstraße 1, 70569 Stuttgart, Germany
[4]Instituto de Física de Líquidos y Sistemas Biológicos (IFLySIB), UNLP-CONICET, 1900 La Plata, Argentina
[5]London Centre for Nanotechnology, University College London, Gower Street, London WC1E 6BT, United Kingdom

*mackenzie@cpfs.mpg.de
§hicks@cpfs.mpg.de

## Abstract

A major area of interest in condensed matter physics is the way electrons in correlated-electron materials can self-organize into ordered states, and a particularly intriguing possibility is that they spontaneously choose a preferred direction of conduction. The correlated-electron metal $Sr_3Ru_2O_7$ has an anomalous phase at low temperatures that features strong susceptibility towards anisotropic transport. This susceptibility has been thought to indicate a spontaneous anisotropy, *i.e.* electronic order that spontaneously breaks the point-group symmetry of the lattice, where external stimuli can then select the orientation of the anisotropy. We investigate further by studying the response of $Sr_3Ru_2O_7$ in the region of phase formation to two fields that lift the native tetragonal symmetry of the lattice: in-plane magnetic field, and orthorhombic lattice distortion through uniaxial pressure. The response to uniaxial pressure is surprisingly strong: compressing the lattice by ~0.1% induces a ~100% transport anisotropy. However, neither the in-plane field nor the pressure phase diagrams are qualitatively consistent with spontaneous symmetry reduction. Instead, both are consistent with a multi-component order parameter that is likely to preserve the point-group symmetry of the lattice, but that is highly susceptible to perturbation.

## Introduction

In 'correlated-electron' metals the interactions between conduction electrons are strong, and at low temperatures can drive spontaneous self-organization into ordered states such as superconductivity and various types of magnetic order. This is analogous to the self-organization of atoms into crystals when a liquid solidifies, but arguably, due to an enhanced role for the effects of quantum mechanics, with a richer range of possibilities. One particularly intriguing possibility is that, solely through the effects of electron-electron interactions, a preferred direction of conduction is established. This occurs when electronic order with two-fold rotation symmetry ($C_2$ symmetry) is established on an otherwise four-fold rotationally symmetric ($C_4$) lattice, as is suspected or demonstrated to occur in cuprate [1, 2] and iron-based superconductors [3, 4], a nematic quantum Hall state in GaAs 2D electron gases [5, 6], the hidden order phase of $URu_2Si_2$ [7], and the superconductivity of $Cu_xBi_2Se_3$ [8]. It has been suspected for a well-known electronic phase in $Sr_3Ru_2O_7$ [9-11].

Apart from cataloguing possible types of order, it is also of broad interest to understand the general conditions under which electronic order can occur. In metals such as copper, the interaction energy is small compared with the zero-point kinetic energy of the charge carriers, and to the lowest temperature so far measured no ordered state is observed. In many cuprates, interactions are very strong, and the resulting ordered phases condense at relatively high temperatures and are robust against perturbation. This robustness is useful for applications but a hindrance for understanding: external stimuli such as laboratory-accessible magnetic fields and pressures do not have much effect. $Sr_3Ru_2O_7$ lies between these extremes. Electronic correlations are strong, as evidenced by a large mass renormalization and Wilson ratio [12, 13], inelastic neutron scattering data [14], and the induction of localized spin density wave order by certain impurities [15, 16]. However electronic order does not appear at ambient pressure and magnetic field, but instead in a narrow window of applied fields near 8 T [10].

Electrical conduction in $Sr_3Ru_2O_7$ takes place mainly in Ru-O bilayers. Interlayer coupling leads to substantial bilayer splitting of parts of the electronic structure, but the bilayers are weakly coupled, and the Fermi surfaces are quasi-two-dimensional [12, 17]. When magnetic field is applied along the $c$ axis (perpendicular to the Ru-O bilayers) a metamagnetic transition, a field-induced jump in the magnetization, occurs at around 8 T. In moderately clean samples, with residual resistivity ~1 µΩ-cm and above, it is a single transition. Metamagnetic transitions are first-order, however the critical endpoint of this transition is located at ~0 K [18], making it a quantum critical endpoint. In samples with residual resistivity below 1 μΩ-cm the metamagnetic transition splits into two first-order transitions, at ≈ 7.9 and ≈ 8.1 T, and an ordered phase, marked by enhanced resistivity, occurs in their vicinity and below 1.1 K [9,10]. In other words, soft modes associated with the quantum critical endpoint appear to nudge the system into an ordered state, and the precise connection between the quantum criticality and condensation of order has been the subject of considerable study [19-23]. This ordered state can be strongly modified by laboratory-accessible magnetic fields and, we will show here, modest uniaxial pressures. This perturbability, along with extremely low disorder – micron-scale mean free paths are achievable – make $Sr_3Ru_2O_7$ an ideal system for study.

Two key observations suggest that the ordered phase spontaneously breaks the tetragonal symmetry of the lattice. One is that the resistivity enhancement in the phase is, in the absence of in-plane magnetic fields, around 100%. This is a strikingly large value and suggests an additional elastic scattering channel, which could be walls between domains of different orientations of the order [10, 24]. The other observation is that within the phase a modest in-plane magnetic field, applied along a ⟨100⟩ lattice direction, induces strong resistive anisotropy between the (100) and (010) directions, suggesting re-orientation of either the order itself or the domain walls [9,11]. (We work throughout this paper with tetragonal notation for $Sr_3Ru_2O_7$, where the ⟨100⟩ directions are the Ru-O-Ru bond directions. $Sr_3Ru_2O_7$ in fact has a slight orthorhombicity, with ⟨110⟩ principle axes [25-27]. However electrical transport in zero field has been found to be isotropic to high precision [43], so for simplicity we refer to $Sr_3Ru_2O_7$ as tetragonal.) Motivated by these observations, the phase has been considered a candidate for nematic order (that is, the point-group symmetry of the lattice is lifted but translation symmetry is preserved), and several models for nematic Fermi surface distortions in $Sr_3Ru_2O_7$ have been proposed [28–32].

A recent landmark neutron scattering study reported the discovery of spin density wave order within the phase: there are scattering peaks at (±$q$,0,0) and (0,±$q$,0), with $q$ incommensurate and taking on slightly different values above and below the second metamagnetic transition (at 8.1 T) [33]. The observed phase boundaries are in excellent agreement with those seen in transport and thermodynamic studies [10]. Most of the models for nematicity in $Sr_3Ru_2O_7$ are based on $q = 0$ inter- and intra-orbital interactions, and while it remains possible that such interactions play an important role in phase formation the neutron data establish that $q \neq 0$ interactions are also present.

Clear understanding of the possible eletronically-induced $C_4$ to $C_2$ symmetry reduction is of broad importance because of the growing range of materials, some listed above, where it is likely to occur. An essential question is to distinguish between a large but finite susceptibility towards $C_2$ symmetry (meaning that small applied symmetry-breaking fields induce strong anisotropy), which in principle is possible with a $C_4$-symmetric ground state, and truly spontaneous $C_2$ symmetry. This can be technically difficult: it is clear from existing knowledge of $Sr_3Ru_2O_7$ that if the anomalous phase region is spontaneously $C_2$-symmetric, it must generally contain domains of both possible orientations [9], and a noninfinitesimal applied field might then be required to fully orient all the domains. But it is important to probe this question precisely, both for general understanding and to constrain microscopic theories for $Sr_3Ru_2O_7$. The neutron study does not address this question because the scattering peaks could arise either from separate domains of (100)- and (010)-oriented spin density waves, or their microscopic coexistence. Thermodynamic measurements do not offer a definitive resolution either: although the transition into the phase with temperature is well-established to be second-order [10, 34, 35], indicating a broken symmetry, it could be the translation symmetry breaking indicated by the neutron data, and not necessarily $C_4$ to $C_2$.

Here, we investigate with two complementary probes that lift the native tetragonal symmetry of the lattice. First, we apply both compressive and tensile in-plane uniaxial pressure to induce orthorhombic lattice distortion. We find that small lattice distortions can dramatically enhance the phase. Then, for gentler and more precise perturbation, we use a vector magnet to study the response to modest, precisely controllable in-plane fields, paying particular attention to the investigation of field-training and hysteretic effects. The results place strong constraints on the energetics of possible domain formation, and yield phase diagrams pointing to a multi-component order parameter that might not break the $C_4$ point-group symmetry of the lattice.

**Results: uniaxial pressure**

It has previously been found that hydrostatic pressure increases the metamagnetic transition fields of $Sr_3Ru_2O_7$ [36, 37], while uniaxial pressure along the *c* axis reduces the transition fields and eventually induces ferromagnetism [38]. In-plane uniaxial pressure experiments have not been reported, although a theoretical study found that nematic transitions may show first-order Griffiths wings in the presence of applied anisotropy [39].

We use a piezoelectric-based apparatus that can both compress and tension test samples [40]. Samples are prepared as beams and secured with epoxy across a gap of width ~1 mm. Piezoelectric actuators strain the sample by minutely varying the width of this gap. A photograph and schematic of a mounted sample are shown in Fig. 1. High strain homogeneity is important to keep transitions reasonably narrow as strain is applied, so samples are prepared with high length-to-width and length-to-thickness aspect ratios, to reduce end effects. Also, the bonding to the sample ends is (approximately) symmetric between the upper and lower surfaces; asymmetry could cause the sample to bend as pressure is applied, imposing a strain gradient across the sample thickness [40].

Because the center of the sample is free, the transverse strains are set by the Poisson's ratios of $Sr_3Ru_2O_7$. Compression along **a**, for example, induces expansion along **b**, so the applied strain induces strong orthorhombic distortion. Both the neutron scattering peaks and transport data indicate $\langle 100 \rangle$ (*i.e.* Ru-O-Ru bond direction) principal axes for the strong $C_2$ susceptibility of the phase: within the phase, an in-plane field along (100) induces strong resistive anisotropy between the (100) and (010) directions, while a (110) field induces very little (110)/(1-10) anisotropy [41]. Therefore, samples are cut (and the pressure applied) along $\langle 100 \rangle$ directions. Three samples were strained, and gave consistent results. They were cut from the same crystal, because our aim was more to test reproducibility in mounting conditions than crystal-to-crystal variation, and the parent crystal has previously been thoroughly characterized with transport, magnetization, and de Haas-van Alphen experiments [42].

Six electrical contacts were made to each sample, as illustrated in Fig. 1A. Longitudinal resistivity $\rho_{aa}$ was measured by running current through contacts 1 and 2 and measuring voltage between contacts 5 and 6, or 3 and 4. The zero-field residual resistivity of the samples was found to be $\approx 0.5$ µΩ-cm. A qualitative measure of the transverse resistivity, $\rho_{bb}$, was obtained for two samples by passing current through contacts 3 and 5 and measuring voltage between contacts 4 and 6. It is only qualitative because (i) the current flow is not homogeneous, so the measured voltage has some dependence on $\rho_{aa}$, and (ii) uncertainty in the true contact geometry prevented accurate calculations to extract $\rho_{bb}$. Also, on one sample, a strong field-dependent but strain-independent background appeared in the transverse response, which was later found to originate in vibration of the transverse leads.

Fig. 1C shows $\rho_{aa}$ against $c$ axis field for a zero-strain reference sample, cut from the same original crystal as the strained samples and mounted on flexible wires to avoid thermal stresses. Also shown is $\rho_{aa}$ of a strained sample, with the strain adjusted to match the reference as closely as possible. Phase formation occurs between 7.9 and $\approx$ 8.6 T. The lower boundary coincides with the first metamagnetic transition. There is a clear resistive anomaly associated with the second metamagnetic transition at $\approx$ 8.1 T, which separates regions of the phase denoted A and B. The existence of the B region is indicated by transport, thermal expansion, and neutron scattering measurements, [33, 35, 43] however the thermodynamic signatures of phase formation are much weaker than for the A region [44].

Fig. 2A shows $\rho_{aa}$ of a strained sample at $T \approx 300$ mK at a series of applied longitudinal strains $\varepsilon_{aa}$. The origin of the strain scale is determined through comparison with the reference, to within an estimated error of $\pm 0.01\%$ strain. (Further discussion of error in the strain is given in the Materials and Methods section.) $\varepsilon_{aa}$ was incremented at 7 T, outside the field range for phase formation. This is important because, if there is spontaneous symmetry breaking then varying $\varepsilon_{aa}$ while within the phase could induce metastable domain configurations. The data show that under compression $\rho_{aa}$ increases strongly and the field range of $\rho_{aa}$ enhancement expands. Under 0.2% compression, phase formation, as identified by enhanced $\rho_{aa}$, extends from 7.7 to 9.3 T. On the other hand, a tension of only 0.05% eliminates the phase-induced enhancement of $\rho_{aa}$.

The transverse response (after background subtraction) is shown in panel B. It increases strongly when the sample is tensioned, which indicates strongly increased $\rho_{bb}$: although the transverse response is also affected by $\rho_{aa}$, under strong tension $\rho_{aa}$ is suppressed to a nearly constant base value. In summary, compression along **a** increases $\rho_{aa}$, and tension along **a** increases $\rho_{bb}$.

Panels A and B of Fig. 3 respectively show contour plots of the longitudinal and transverse responses; contour plots for another strained sample are shown in Supplemental Material. The longitudinal and transverse responses approximately mirror each other, and for each there is a region of strongly enhanced response, that indicates phase formation. Taking a fixed contour level as an approximate criterion for the boundaries of these regions yields the phase diagram shown in panel C. A notable feature of Fig. 3C is the existence of a range of overlap, with a maximum width of ≈ 0.08% strain. This overlap suggests two components of the order that can be simultaneously nonzero, so it is evidence for a multi-component order parameter. We discuss more thoroughly in the Discussion.

We conclude our presentation of strain-tuning data by showing, in Fig. 4A, the effect of strain over a wider field range. Outside the phase formation region, the strain response is nearly linear over the applied strain range, so plotted in the figure is the normalized linear elastoresistance, $(1/R)dR/d\varepsilon_{aa}$, determined at $\varepsilon_{aa} = 0$. The change in sample dimensions yields a geometric contribution to the elastoresistance even when the resistivity does not change. For isotropic materials with Poisson's ratio $\nu$ this geometric contribution is $1 + 2\nu$, typically ~ 2, and for ordinary metals it is typically the dominant contribution [45]. In narrow band metals the change in hopping can substantially enhance the elastoresistance, as can electronic correlations [4]. $Sr_3Ru_2O_7$ is known to be both narrow-band and strongly correlated, which is consistent with the observation of $(1/R)dR/d\varepsilon_{aa}$ ~ 20 far from the phase: a much smaller elastoresistance than within the phase, but still large. Also notable is that the elastoresistance is nearly zero at 7.6 and 9.2 T, a feature that appears as crossing points in Fig. 2A. The data in Fig. 4B show that at these fields $\rho_{aa}$ is nearly strain-independent for $|\varepsilon_{aa}|$ up to ~ 0.1%. This appears to reflect a balance between the positive elastoresistance outside the phase and the strongly negative elastoresistance within, and it will be interesting in future to examine these points more carefully.

**Results: vector field**

Another way to lift $C_4$ symmetry is with a vector magnetic field, $\mathbf{H} = (H_a, H_b, H_c)$. Recently-published vector field results showed that the high susceptibility to in-plane fields of the A phase is shared by the B phase [43]. The vector magnet used in both that and the present work can apply $H_c$ to 8.5 T, simultaneously with $\sqrt{(H_a^2+H_b^2)}$ up to 0.9 T, allowing the field angle to be swept across a range $\approx \arctan(0.9/8.5) - \arctan(-0.9/8.5) = 12°$. Samples were mounted on a mechanical rotator to allow the sample $c$ axis to be shifted relative to this accessible range.

Three samples were measured in the vector magnet: a high-aspect-ratio sample optimized for sensitivity to $\rho_{aa}$, and two octagonal samples. The octagonal shape provides shape symmetry between the $a$ and $b$ axes, and also between $\langle 100 \rangle$ and $\langle 110 \rangle$ directions, to avoid biasing possible domain structures [46]. All samples were mounted on flexible wires to avoid strain effects. We present data from the octagons, although our conclusions are supported by all three samples independently. The contact configuration for the octagonal samples is illustrated in Fig. 5. An $a$-axis resistance is measured by applying a current through contacts 1 and 5 and measuring the voltage between contacts 2 and 4, or 8 and 6. Rotating the current and voltage contacts by 90° yields a $b$-axis resistance. With this geometry the current flow is not perfectly homogeneous, so the measured $a$-axis resistance has some dependence on $\rho_{bb}$, and vice versa, however the shape symmetry of the sample makes it unambiguous whether a prominent feature is due primarily to a change in $\rho_{aa}$ or $\rho_{bb}$.

We start with a field rotation study: $R_{aa} = (V_2 - V_4)/I_{15}$ and $R_{bb} = (V_4 - V_6)/I_{37}$ were measured while rotating $\mathbf{H}$ about $\mathbf{c}$ at a fixed polar angle $\theta = 6°$. Results are shown in

Fig. 5. Both quantities are found to vary smoothly, and approximately as cos(2$\phi$). Field rotation studies on the high-aspect-ratio sample, for $\theta$ between 1° and 10.5°, similarly showed a smooth dependence of $\rho_{aa}$ on $\phi$, approximately proportional to cos(2$\phi$) for $\theta$ below ~ 8°.

This smooth dependence of $R_{aa}$ and $R_{bb}$ on $\phi$ does not match naïve expectations for an Ising-type $C_2$-symmetric order parameter, for which sharp steps are expected when the orientation of the order flips. However, if order of this type is broken up into domains then the reorientation of individual domains may yield a series of smaller steps, known in ferromagnets as Barkhausen jumps, rather than a single large step. If domains are very small, individual steps might not be visible at all, but thermal or quantum fluctuations of the domain orientation may yield observable excess noise such as that reported in $YBa_2Cu_3O_{6+x}$ [47, 48]. Our measurements on $Sr_3Ru_2O_7$ resolved neither Barkhausen-like jumps nor excess noise. This is notable evidence against an Ising-type $C_2$-symmetric order, especially as the very low disorder of the sample means that a fine-scale domain structure is not generally expected. However on its own this evidence is not conclusive because it might be an issue of experimental resolution.

The cos(2$\phi$) form of $R_{aa}$ and $R_{bb}$ is consistent with a quadratic coupling to in-plane field. In the absence of in-plane ferromagnetism the response to in-plane field should be symmetric with respect to the sign of the field, so for small in-plane fields where a lowest-order expansion applies one expects generally $R_{aa} = R_{aa,0} + aH_a^2 + bH_b^2 = R_{aa,0} + H^2(a\cos^2\phi + b\sin^2\phi)\cos^2\theta$, where $R_{aa,0}$ is the resistance with $H_a = H_b = 0$, and $a$ and $b$ are coupling constants. A cos 2$\phi$ variation of $R_{aa}$ results when $a \neq b$.

To complement the strain data shown in Fig. 3, we mapped $R_{aa}$ and $R_{bb}$ against $H_a$ and $H_c$ (keeping $H_b = 0$). $R_{aa}$ was measured while sweeping |**H**| through the phase at fixed $\theta$, then $R_{bb}$ in a second |**H**| sweep at the same $\theta$. $\theta$ was incremented outside the field range of phase formation, to minimize the chances of creating metastable domain structures in case of a spontaneously $C_2$-symmetric order. Results of measurements at nine nonzero values of $\theta$ are shown in Fig. 6, and contour plots of $R_{aa}$ and $R_{bb}$ are presented in Fig. 7A. The vector field and strain measurements are complementary because although achievable strains can perturb the phase much more strongly than in-plane field, the transitions are sharper in the vector field data, so the overlap region identified in the strain data might be resolved more clearly in the vector field data. The transitions can be readily identified by changes in slope of $R$ against $|\mathbf{H}|$, as illustrated in Fig. 6. On the contour plots shown in Fig. 7A, the transitions are indicated by black points, linked as a guide to the eye by black lines. Following the quadratic dependence on in-plane field found in Fig. 5, the data are plotted against $H_a^2$. The regions of enhanced $R_{aa}$ and $R_{bb}$, bounded by the transitions, are shown together in panel B. As in the strain phase diagram (Fig. 3), there is a region of overlap where both $R_{aa}$ and $R_{bb}$ are enhanced, that extends out to $(\mu_0 H_a)^2 \approx (1.7\text{ T})^2$.

If the large susceptibility of the phase to in-plane field is due to domain movement, hysteresis might be expected when the applied field is varied. The high precision and absence of sample heating make the vector magnet an ideal tool to search for this hysteresis. Results of a test in which $R_{bb}$ was measured while increasing then decreasing $H_a$ are shown in Fig. 8. $H_c$ was simultaneously controlled to stay within the overlap region, up to $\mu_0 H_a \approx 1.5$ T: if the simultaneous enhancement of $\rho_{aa}$ and $\rho_{bb}$ within the overlap region is due to coexisting domains of (100)- and (010)-oriented orders, exiting and re-entering the phase with the strongest possible in-plane field should give the largest possible change in domain configuration. The data show no resolvable hysteresis, however. This test was performed at 50 mK, a factor of twenty lower than the transition temperature into the phase. Another test for hysteresis consisted of in-plane field rotations, like that illustrated in Fig. 5, with both clockwise and counter-clockwise rotation. The rotations were performed at values of $\theta$ between 6° and 14°, and spanned $\phi = 45°$, where, for spontaneous $C_2$ symmetry, the applied field should start to flip domains between (100) and (010) orientations. Again, no definite hysteresis was observed.

**Discussion**

The measurements described above have confirmed the strong susceptibility of magnetotransport in $Sr_3Ru_2O_7$ to two $C_2$-symmetric fields, orthorhombic lattice distortion through uniaxial pressure and in-plane magnetic field. The former is a probe that has not previously been applied to $Sr_3Ru_2O_7$, while the vector magnetic field experiments described above have probed for hysteretic and noise signatures of domains with much higher precision than in any previous work.

In this Discussion, we first consider more carefully whether our results could be explained by domains of a $C_2$-symmetric order. We then briefly discuss possible forms of microscopic coexistence of density waves relevant to $Sr_3Ru_2O_7$ that could explain our data, and close by discussing challenges in explaining the resistivity of the anomalous phase solely with density wave order.

We believe that our results present a considerable challenge to models that appeal to the formation and movement of domains to explain the transport properties of $Sr_3Ru_2O_7$, that goes beyond the observed lack of hysteresis. The basic issue is the following. In the experiments we repeatedly entered and exited the phase in the presence of non-zero $C_2$ field, so that any domain configuration should be at or near the ground state under the applied $C_2$ field. For domains to account for the resistive properties of $Sr_3Ru_2O_7$, they must persist to at least the upper limits of the overlap regions. As shown in Figs. 3c) and 7b), this means that domains should exist, in the ground state, in strains up to $\sim 4 \cdot 10^{-4}$ and in-plane fields up to $\sim 1.7$ T. For this to be the case, there must be a free energy minimum for domain formation that is stronger than an interaction energy related to symmetry-breaking fields of these strengths. Examination of simple candidate mechanisms illustrates the difficulty with this approach.

One possible mechanism for ground-state domain formation is long-range interactions, which may be elastic or magnetic. If $C_2$-symmetric electronic order induces $C_2$ lattice deformation that exceeds the applied average strain, domain formation would reduce the total elastic energy of the system. The domain walls would be closely analogous to twin boundaries of an orthorhombic lattice [24]. However this mechanism is not quantitatively consistent with observations. The lattice distortion resulting from in-plane fields high enough to saturate the resistive anisotropy is known to be $\sim 4 \cdot 10^{-6}$ [49]. That is, if the phase is spontaneously $C_2$ symmetric, elastic interactions would favour domain formation only up to an applied average strain of $\sim 4 \cdot 10^{-6}$, two orders of magnitude less than the observed extent of the overlap region (up to strains of $\sim 4 \cdot 10^{-4}$). Similarly, long-range magnetic interactions could favor domain formation if the domains have differing magnetizations [50]. However, the order would need to include in-plane ferromagnetism on at least a ~1.7 T scale to stabilize domains against a 1.7 T applied in-plane field, and there is no expectation of that. For comparison, the larger metamagnetic jump, at 7.9 T, has a magnitude of $\mu_0 \Delta M = 0.008$ T [34].

Another possible mechanism is random-field disorder, which could result *e.g.* from locally-oriented defects or inhomogeneous strain. A $C_2$-symmetric order might have domains in the ground state if the internal field has spatially-varying orientation. In this case, as the in-plane field is ramped the ground-state domain distribution should in general transition smoothly from all-(100) to all-(010), with the width of the transition determined by the distribution of internal fields. Hysteresis and glassy dynamics are also possible, especially if the disorder has a short length scale [48]. However, in both the strain and vector-field data the transitions into the overlap regions are sharp compared to the width of the overlap, against expectations for strong disorder broadening. This is seen particularly clearly in Fig. 8, where the form of $R_{aa}(H_a)$ is much more consistent with a transition at $(\mu_0 H_a)^2 \approx 2.4$ T$^2$, than with disorder broadening of a (100)-to-(010) transition at $H_a=0$.

The above analysis is based on specific mechanisms for domain formation, but the central issue is more general: any domains present in the system have to be stable against substantial $C_2$ fields and yet also be associated with extremely weak hysteresis, below the resolution of our measurements. In general these are contradictory requirements.

Given the rather strong arguments against the presence of domains in the anomalous state, the most natural conclusion is that the observed resistivity in the phase is its intrinsic resistivity, not a result of domain wall scattering. Further, our data and analysis argue for the intrinsic order being multi-component, with microscopic coexistence of (100)- and (010)-oriented components when applied $C_2$ fields are not too large. The argument that the resistivity is an intrinsic property of the phase is supported by the fact that strong resistive anisotropy, and strong resistivity enhancement along the "hard" direction, persists even under strong $C_2$ fields, where samples have been argued to be mono-domain [33,49].

With the above conclusions in mind, the close correlation observed in the neutron scattering data of Ref. [33] between spin density wave order and resistivity in the anomalous phase means that the simplest hypothesis is to associate the two components with spin density waves. We cannot rule out the possibility that the spin modulations observed in the neutron experiments are connected with further, as-yet unobserved, charge order, but in the absence of concrete observations in the charge sector any discussion of this would be speculative.

In the scenario that the resistivity is directly linked to spin density wave formation, there are several possibilities for microscopic coexistence. The simplest is that the (100)- and (010)-oriented waves coexist within each layer, yielding a checkerboard-type order, perhaps similar to the tetragonal spin density wave state recently discovered in iron pnictide superconductors [51]. The coexistence could also be microscopic but spatially separated, with the (100) and (010) components existing in alternate layers within each bilayer, yielding an inversion-symmetry-broken state. This would be similar to the intralayer nematic order discussed in Refs. 52-54, where it is proposed that $Sr_3Ru_2O_7$ could host nematic order but with the orientation of the nematicity alternating from layer to layer. Another possibility is that the (100) and (010) components exist in alternating bilayers.

Although we believe that the evidence for microscopic coexistence is strong, the high susceptibility of the phase to $C_4$-symmetry-breaking fields is surprising. It may reflect fine-tuning, in that the order is only weakly stable in the unstrained lattice and hence has a high general susceptibility to perturbation. However, our observations could also reflect a more specific high susceptibility towards $C_2$ symmetry, for example if the (100) and (010) components compete reasonably strongly but not strongly enough to prevent microscopic coexistence. Another possibility is that, if the (100) and (010) components alternate between layers or bilayers, the coupling stabilizing this alternation is likely to be weak, such that the density waves might re-orient with modest applied $C_2$ fields. In general, it would be interesting to determine if there is a deeper reason for the high perturbability of the phase. We also note that the strong enhancement of the phase with strain could have bearing on studies of the links between phase formation and metamagnetic quantum criticality: under strains above

~0.1% the region of phase formation appears to extend to fields considerably above the metamagnetic transitions. We believe this is a point deserving further study.

Finally, we note that although our data are qualitatively inconsistent with a domain scenario, an outstanding challenge to any model not relying on strong domain wall scattering is to understand the magnitude of the resistivity enhancement. We have proposed direct identification of $\rho_{aa}$ and $\rho_{bb}$ enhancements with magnitudes of (100)- and (010)-oriented order parameter components, that is, the intrinsic resistivity of the ordered phase without observable domain wall scattering. This model provides a qualitatively simple explanation for our data. However, it is not easy to understand the scale of the resistive change. At large strains, the ratio of resistivities parallel and perpendicular to the density wave $\rho_\parallel/\rho_\text{perp}$ is ~ 2.4. For comparison, in elemental chromium, a three-dimensional metallic spin density wave system with $T_N = 312$ K, $\rho_\parallel/\rho_\text{perp}$ saturates at only 1.08 [55]. $Sr_3Ru_2O_7$ has at least seven Fermi surface sheets [17] and it does not appear possible that nesting across a single wave vector could gap away enough carriers to yield such strong anisotropy. Widespread gapping also appears inconsistent with observed thermodynamic properties: the entropy and specific heat are highest in the region of phase formation [34, 56, 57], indicating a large number of ungapped carriers. If, as the experiments reported here suggest, the observed resistivity is intrinsic and not due to domain wall scattering, understanding the magnitude of the changes in resistivity is a key issue for future investigations of $Sr_3Ru_2O_7$.

**Materials and Methods**

Samples were grown by a floating zone method, [58] and cut to the desired shape with a wire saw. The strained samples were 40–45 μm thick and 170–280 μm wide. Electrical contacts were made with DuPont 6838 silver paste, cured at 450° C for ≈5 minutes. The strained samples were secured in the strain apparatus with Stycast 2850, cured at a temperature of ≈65° for several hours to obtain a harder cure than at room temperature. For strained samples #1, #2, and #3, the epoxy thickness was ≈80, 40, and 20 μm, respectively. For low-noise measurements of resistivity, on strained sample #1 and for the vector field data we used low-temperature transformers to amplify the signal voltage. For samples #2 and #3, room-temperature transformers were used.

Strained sample #1 was mounted in the same strain apparatus reported in Refs. [40] and [59], which used a strain gauge for measurement of the displacement applied to the sample. Samples #2 and #3 were mounted in a newer apparatus with a capacitive displacement sensor. Through comparison of the elastoresistive response outside the region of phase formation, it was found that strains reported from the original apparatus needed to be multiplied by a factor of 1.6 to match those from the newer apparatus, suggesting that the strain gauge somewhat impeded the motion of the original apparatus. This correction factor was verified with measurements on $Sr_2RuO_4$, and is included in strains reported for sample #1. All strains reported in this paper are also corrected for an estimate of the epoxy deformation: through finite element calculations we estimate that 52%, 68%, and 78%, respectively, of the applied displacement was transmitted as sample strain to the exposed portion of samples #1, #2, and #3. In these calculations we took the Young's modulus of the epoxy to be 15 GPa, and for $Sr_3Ru_2O_7$ we took the elastic tensor of $Sr_2RuO_4$. [60] Uncertainty in dimensions and elastic properties gives a ~20% proportional error on all strains quoted in this work. No hysteresis against applied displacement was observed, indicating that the epoxy deformation was always elastic.

Due to field dependence of the temperature sensor, the temperature of the strained samples varied by ≈20 mK between 7 and 10 T.

**References**

[1] Y. Ando, K. Segawa, S. Komiya, and A.N. Lavrov. Electrical Resistivity Anisotropy from Self-Organized One Dimensionality in High-Temperature Superconductors. *Phys. Rev. Lett.* **88** 137005 (2002).

[2] V. Hinkov, D. Haug, B. Fauqué, P. Bourges, Y. Sidis, A. Ivanov, C. Bernhard, C.T. Lin, B. Keimer. Electronic Liquid Crystal State in the High-Temperature Superconductor $YBa_2Cu_3O_{6.45}$. *Science* **319** 597 (2008).

[3] J.-H. Chu, J.G. Analytis, K. De Greeve, P.L. McMahon, Z. Islam, Y. Yamamoto, and I.R. Fisher. In-Plane Resistivity Anisotropy in an Underdoped Iron Arsenide Superconductor. *Science* **329** 824 (2010).

[4] J.-H. Chu, H.-H. Kuo, J.G. Analytis, and I.R. Fisher. Divergent Nematic Susceptibility in an Iron Arsenide Superconductor. *Science* **337** 710 (2012).

[5] J. Pollanen, K.B. Cooper, S. Brandsen, J.P. Eisenstein, L.N. Pfeiffer, and K.W. West. Heterostructure symmetry and the orientation of the quantum Hall nematic phases. *Phys. Rev. B* **92** 115410 (2015).

[6] M.P. Lilly, K.B. Cooper, J.P. Eisenstein, L.N. Pfeiffer, and K.W. West. Evidence for an Anisotropic State of Two-Dimensional Electrons in High Landau Levels. *Phys. Rev. Lett.* **82** 394 (1999).

[7] R. Okazaki, T. Shibauchi, H.J. Shi, Y. Haga, T.D. Matsuda, E. Yamamoto, Y. Onuki, H. Ikeda, and Y. Matsuda. Rotational Symmetry Breaking in the Hidden-Order Phase of $URu_2Si_2$. *Science* **331** 439 (2011).

[8] S. Yonezawa, K. Tajiri, S. Nakata, Y. Nagai, Z. Wang, K. Segawa, Y. Ando, and Y. Maeno. Thermodynamic evidence for nematic superconductivity in $Cu_xBi_2Se_3$. *Nature Physics* doi:10.1038/nphys3907 (2016).

[9] A.P. Mackenzie, J.A.N. Bruin, R.A. Borzi, A.W. Rost, and S.A. Grigera. Quantum criticality and the formation of a putative electronic liquid crystal in $Sr_3Ru_2O_7$. *Physica C* **481** 207 (2012).

[10] S.A. Grigera, P. Gegenwart, R.A. Borzi, F. Weickert, A.J. Scho_eld, R.S. Perry, T. Tayama, T. Sakakibara, Y. Maeno, A.G. Green, A.P. Mackenzie. Disorder-Sensitive Phase Formation Linked to Metamagnetic Quantum Criticality. *Science* **306** 1154 (2004).

[11] R.A. Borzi, S.A. Grigera, J. Farrell, R.S. Perry, S.J.S. Lister, S.L. Lee, D.A. Tennant, Y. Maeno, and A.P. Mackenzie. Formation of a Nematic Fluid at High Fields in $Sr_3Ru_2O_7$. *Science* **315** 214 (2007).

 [12] M.P. Allan, A. Tamai, E. Rozbicki, M.H. Fischer, J. Voss, P.D.C. King, W. Meevasana, S. Thirupathaiah, E. Rienks, J. Fink, D.A. Tennant, R.S. Perry, J.F. Mercure, M.A. Wang, Jinho Lee, C.J. Fennie, E.-A. Kim, M.J. Lawler, K.M. Shen, A.P. Mackenzie, Z.-X. Shen, and F. Baumberger. Formation of heavy *d*-electron quasiparticles in $Sr_3Ru_2O_7$. *New Journal of Physics* **15** 063029 (2013).

[13] S.-I. Ikeda, Y. Maeno, S. Nakatsuji, M. Kosaka, and Y. Uwatoko. Ground state in $Sr_3Ru_2O_7$: Fermi liquid close to a ferromagnetic instability. *Phys. Rev. B* **62** R6089 (2000).

[14] L. Capogna, E.M. Forgan, S.M. Hayden, A. Wildes, J.A. Du_y, A.P. Mackenzie, R.S. Perry, S. Ikeda, Y. Maeno, and S.P. Brown. Observation of two-dimensional spin fluctuations in the bilayer ruthenate $Sr_3Ru_2O_7$ by inelastic neutron scattering. *Phys. Rev. B* **62** 012504 (2003).

[15] R. Mathieu, A. Asamitsu, Y. Kaneko, J.P. He, X.Z. Yu, R. Kumai, Y. Onose, N. Takeshita, T. Arima, H. Takagi, and Y. Tokura. Impurity-induced transition to a Mott insulator in $Sr_3Ru_2O_7$. *Phys. Rev. B* **72** 092404 (2005).

[16] M.A. Hossain, B. Bohnenbuck, Y.D. Chuang, M.W. Haverkort, I.S. Elfimov, A. Tanaka, A.G. Cruz Gonzalez, Z. Hu, H.-J. Lin, C.T. Chen, R. Mathieu, Y. Tokura, Y. Yoshida, L.H. Tjeng, Z. Hussain, B. Keimer, G.A. Sawatzky, and A. Damascelli. Mott versus Slater-type metalinsulator transition in Mn-substituted $Sr_3Ru_2O_7$. *Phys. Rev. B* **86** 041102 (2012).

[17] A. Tamai, M.P. Allan, J.F. Mercure, W. Meevasana, R. Dunkel, D.H. Lu, R.S. Perry, A.P. Mackenzie, D.J. Singh, Z.-X. Shen, and F. Baumberger. Fermi Surface and van Hove Singularities in the Itinerant Metamagnet $Sr_3Ru_2O_7$. *Phys. Rev. Lett.* **101** 026407 (2008).

[18] S.A. Grigera, R.A. Borzi, A.P. Mackenzie, S.R. Julian, R.S. Perry, and Y. Maeno. Angular dependence of the magnetic susceptibility in the itinerant metamagnet $Sr_3Ru_2O_7$.

*Phys. Rev. B* **67** 214427 (2003).

[19] A.G. Green, S.A. Grigera, R.A. Borzi, A.P. Mackenzie, R.S. Perry, and B.D. Simons. Phase Bifurcation and Quantum Fluctuations in $Sr_3Ru_2O_7$. *Phys. Rev. Lett.* **95** 086402 (2005).

[20] H.-Y. Kee and Y.B. Kim. Itinerant metamagnetism induced by electronic nematic order. *Phys. Rev. B* **71** 184402 (2005).

[21] H. Yamase. Mean-field theory on a coupled system of ferromagnetism and electronic nematic order. *Phys. Rev. B* **87** 195117 (2013).

[22] A.M. Berridge, A.G. Green, S.A. Grigera, and B.D. Simons. Inhomogeneous Magnetic Phases: A Fulde-Ferrell-Larkin-Ovchinnikov-Like Phase in $Sr_3Ru_2O_7$. *Phys. Rev. Lett.* **102** 136404 (2009).

[23] A.M. Berridge, S.A. Grigera, B.D. Simons, and A.G. Green. Magnetic analog of the Fulde-Ferrell-Larkin-Ovchinnikov phase in $Sr_3Ru_2O_7$. *Phys. Rev. B* **81** 054429 (2010).

[24] H.-J. Doh, Y.B. Kim, and K.H. Ahn. Nematic Domains and Resistivity in an Itinerant Metamagnet Coupled to a Lattice. *Phys. Rev. Lett.* **98** 126407 (2007).

[25] H. Shaked, J.D. Jorgensen, O. Chmaissem, S. Ikeda, and Y. Maeno. Neutron Di_raction Study of the Structural Distortions in $Sr_3Ru_2O_7$. *J. Solid State Chemistry* **154** 361 (2000).

[26] R Kiyanagi, K Tsuda, N Aso, H Kimura, Y Noda, Y Yoshida, S Ikeda, and Y Uwatoko. Investigation of the Structure of Single Crystal $Sr_3Ru_2O_7$ by Neutron and Convergent Beam Electron Diffractions. *J. Phys. Soc. Japan* **73** 639 (2004).

[27] Biao Hu, G.T. McCandless, M. Menard, V.B. Nascimento, J.Y. Chen, E.W. Plummer, and R. Jin. Surface and bulk structural properties of single-crystalline $Sr_3Ru_2O_7$. *Phys. Rev. B* **81** 184104 (2010).

[28] H. Yamase and A.A. Katanin. Van Hove Singularity and Spontaneous Fermi Surface Symmetry Breaking in $Sr_3Ru_2O_7$. *J. Phys. Soc. Japan* **76** 073706 (2007).

[29] W.-C. Lee and C.-J. Wu. Theory of unconventional metamagnetic electron states in orbital band systems. *Phys. Rev. B* **80** 104438 (2009).

[30] S. Raghu, A. Paramekanti, E.A. Kim, R.A. Borzi, S.A. Grigera, A.P. Mackenzie, and S.A. Kivelson. Microscopic theory of the nematic phase in $Sr_3Ru_2O_7$. *Phys. Rev. B* **79** 214402 (2009).

[31] C.M. Puetter, J.G. Rau, and H.-Y. Kee. Microscopic route to nematicity in $Sr_3Ru_2O_7$. *Phys. Rev. B* **81** 081105(R) (2010).

[32] M. Tsuchiizu, Y. Ohno, S. Onari, and H. Kontani. Orbital Nematic Instability in the Two- Orbital Hubbard Model: Renormalization-Group + Constrained RPA Analysis. *Phys. Rev. Lett.* **111** 057003 (2013).

[33] C. Lester, S. Ramos, R.S. Perry, T.P. Croft, R.I. Bewley, T. Guidi, P. Manueal, D.D. Khalyavin, E.M. Forgan, and S.M. Hayden. Field-tunable spin-density-wave phases in $Sr_3Ru_2O_7$. *Nature Materials* **14** 373 (2015).

[34] A.W. Rost, R.S. Perry, J.F. Mercure, A.P. Mackenzie, and S.A. Grigera. Entropy Landscape of Phase Formation Associated with Quantum Criticality in $Sr_3Ru_2O_7$. *Science* **325**

1360 (2009).

[35] C. Stingl, R.S. Perry, Y. Maeno, and P. Gegenwart. Electronic nematicity and its relation to quantum criticality in $Sr_3Ru_2O_7$ studied by thermal expansion. *Physica Status Solidi B* **250** 450 (2013).

[36] W. Wu, A. McCollam, S.A. Grigera, R.S. Perry, A.P. Mackenzie, and S.R. Julian. Quantum critical metamagnetism of $Sr_3Ru_2O_7$ under hydrostatic pressure. *Phys. Rev. B* **83** 045106 (2011).

[37] M. Chiao, C. Pfleiderer, S.R. Julian, G.G. Lonzarich, R.S. Perry, A.P. Mackenzie, Y. Maeno. Effect of pressure on metamagnetic $Sr_3Ru_2O_7$. *Physica B* **312-313** 698 (2002).

[38] S.-I. Ikeda, N. Shirakawa, T. Yanagisawa, Y. Yoshida, S. Koikegami, S. Koike, M. Kosaka, and Y. Uwatoko. Uniaxial-Pressure Induced Ferromagnetism of Enhanced Paramagnetic $Sr_3Ru_2O_7$. *J. Phys. Soc. Japan* **73** 1322 (2004).

[39] H. Yamase. Electronic nematic phase transition in the presence of anisotropy. *Phys. Rev. B* **91** 195121 (2015).

[40] C.W. Hicks, M.E. Barber, S.D. Edkins, D.O. Brodsky, and A.P. Mackenzie. Piezoelectric-based apparatus for strain-tuning. *Rev. Sci. Inst.* **85** 065003 (2014).

[41] J.A.N. Bruin, R. Borzi, S.A. Grigera, R.S. Perry, and A.P. Mackenzie. *In preparation*.

[42] J.-F. Mercure, A.W. Rost, E.C.T. O_Farrell, S.K. Goh, R.S. Perry, M.L. Sutherland, S.A. Grigera, R.A. Borzi, P. Gegenwart, A.S. Gibbs, and A.P. Mackenzie. Quantum oscillations near the metamagnetic transition in $Sr_3Ru_2O_7$. *Phys. Rev. B* **81** 235103 (2010).

[43] J.A.N. Bruin, R.A. Borzi, S.A. Grigera, A.W. Rost, R.S. Perry, and A.P. Mackenzie. Study of the electronic nematic phase of $Sr_3Ru_2O_7$ with precise control of the applied magnetic field vector. *Phys. Rev. B* **87** 161106(R) (2013).

[44] D. Sun, A.W. Rost, R.S. Perry, A.P. Mackenzie, and M. Brando arXiv:1605.00396

[45] H.-H. Kuo, M.C. Shapiro, S.C. Riggs, and I.R. Fisher. Measurement of the elastoresistive coefficients of the underdoped iron arsenide $Ba(Fe_{0.975}Ca_{0.025})_2As_2$. *Phys. Rev. B* **88** 085113 (2013).

[46] R.A. Borzi, A. McCollam, J.A.N. Bruin, R.S. Perry, A.P. Mackenzie, and S.A. Grigera. Hall coefficient anomaly in the low-temperature high-field phase of $Sr_3Ru_2O_7$. *Phys. Rev. B* **84** 205112 (2011).

[47] D.S. Caplan, V. Orlyanchik, M.B. Weissman, D.J. Van Harlingen, E.H. Fradkin, M.J. Hinton, and T.R. Lemberger. Anomalous Noise in the Pseudogap Regime in $YBa_2Cu_3O_{7-\delta}$. *Phys. Rev. Lett.* **104** 177001 (2010).

[48] E.W. Carlson, K.A. Dahmen, E. Fradkin, and S.A. Kivelson. Hysteresis and Noise from Electronic Nematicity in High-Temperature Superconductors. *Phys. Rev. Lett.* **96** 097003 (2006).

[49] C. Stingl, R.S. Perry, Y. Maeno, and P. Gegenwart. Symmetry-Breaking Lattice Distortion in $Sr_3Ru_2O_7$. *Phys. Rev. Lett.* **107** 026404 (2011).

[50] B. Binz, H.B. Braun, T.M. Rice, M. Sigrist. Magnetic Domain Formation in Itinerant

Metamagnets. *Phys. Rev. Lett.* **96** 196406 (2006).

[51] Xiaoyu Wang and R.M. Fernandes. Impact of local-moment fluctuations on the magnetic degeneracy of iron arsenide superconductors. *Phys. Rev. B* **89** 144502 (2014).

[52] C. Puetter, H. Doh, and H.-Y. Kee. Metanematic transitions in a bilayer system: Application to the bilayer ruthenate. *Phys. Rev. B* **76** 235112 (2007).

[53] H. Yamase. Spontaneous Fermi surface symmetry breaking in bilayer systems. *Phys. Rev. B* **80** 115102 (2009).

[54] T. Hitomi and Y. Yanase. Electric Octupole Order in Bilayer Ruthenate $Sr_3Ru_2O_7$. *J. Phys. Soc. Japan* **83** 114704 (2014).

[55] W.B. Muir and J.O. Ström-Olsen. Electrical Resistance if Single-Crystal Single-Domain Chromium from 77 to 325° K. *Phys. Rev. B* **4** 988 (1971).

[56] P. Gegenwart, F. Weickert, M. Garst, R.S. Perry, and Y. Maeno. Metamagnetic Quantum Criticality in $Sr_3Ru_2O_7$ Studied by Thermal Expansion. *Phys. Rev. Lett.* **96** 136402 (2006).

[57] A.W. Rost, S.A. Grigera, J.A.N. Bruin, R.S. Perry, D. Tian, S. Raghu, S.A. Kivelsone, and A.P. Mackenzie. Thermodynamics of phase formation in the quantum critical metal $Sr_3Ru_2O_7$. *Proc. Nat. Acad. Sciences* **108** 16549 (2011).

[58] R. S. Perry and Y. Maeno. Systematic approach to the growth of high-quality single crystals of $Sr_3Ru_2O_7$. *J. Cryst. Growth* **271** 134 (2004).

[59] C.W. Hicks, D.O. Brodsky, E.A. Yelland, A.S. Gibbs, J.A.N. Bruin, M.E. Barber, S.D. Edkins, K. Nishimura, S. Yonezawa, Y. Maeno, and A.P. Mackenzie. Strong Increase of $T_c$ of $Sr_2RuO_4$ Under Both Compressive and Tensile Strain. *Science* **344** 283 (2014).

[60] J.P. Paglione, C. Lupien, W.A. MacFarlane, J.M. Perz, L. Taillefer, Z.Q. Mao, and Y. Maeno. Elastic tensor of $Sr_2RuO_4$. *Phys. Rev. B* **65** 220506 (2002).

**Acknowledgements:** We acknowledge useful discussions with J.J. Betouras, E. Berg, R.M. Fernandes, A.G. Green, S.M. Hayden, C.A. Hooley, S.A. Kivelson, C. Lester, and W. Metzner. **Funding:** This work was supported by the Max Planck Society, the Engineering and Physical Sciences Research Council, UK (grants EP/1031014/1 and EP/G03673X/1), Agencia Nacional de Promoción Científica y Tecnológica, Argentina, through PICT 2013 N•2004 and PICT N•2618, and Consejo Nacional de Investigaciones Científicas y Técnicas. **Author contributions:** D.B. and M.B. carried out measurements on strain-tuning of $Sr_3Ru_2O_7$. J.B., R.B, and S.G. carried out the vector field measurements. R.P. grew the samples. A.M. and C.H. devised and supervised the project. The manuscript was drafted by C.H., A.M., D.B., R.B., S.G., and J.B. **Competing Interests:** C.H. has 31% ownership of Razorbill Instruments, which has commercialized apparatus based on that used in this work. **Data and materials availability:** Data underpinning this publication can be accessed at ***.

**Supplementary Materials:** The supplementary materials include Figure S1, which shows data for strained sample #2.

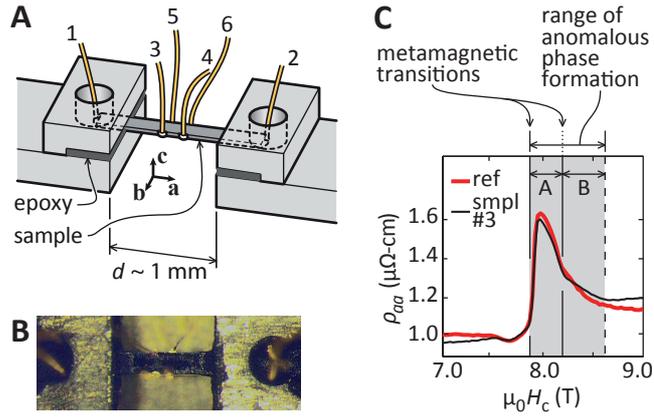

Figure 1: **Sample setup for strain tuning.** **A** and **B** Schematic illustration and photograph of a mounted sample. Strain is applied by minutely varying the gap width $d$. **C**. $\rho_{aa}$ against $H_c$, where $\mathbf{H} = (0, 0, H_c)$, of a zero-strain reference sample, and strained sample #3 mounted in the strain apparatus and tuned to near-zero strain. T ≈ 300 mK.

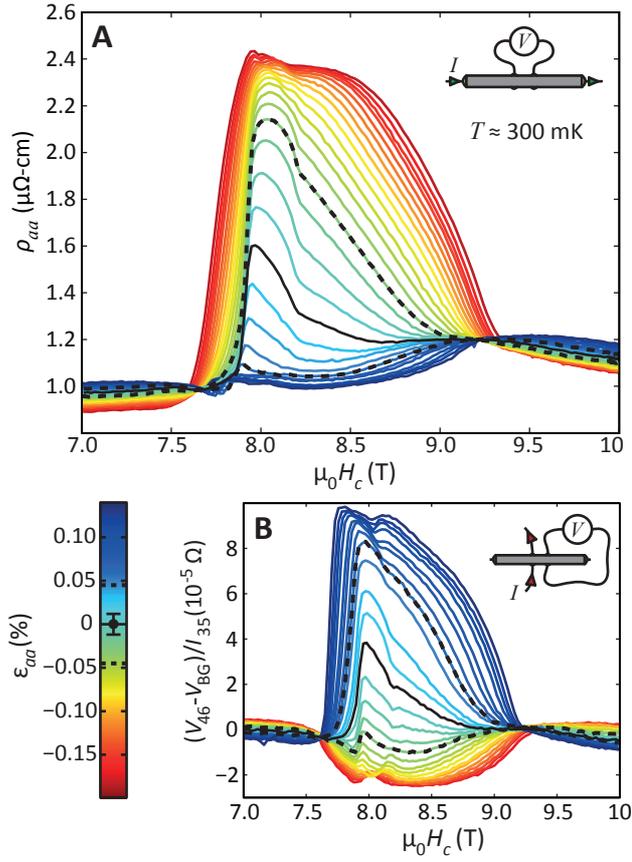

Figure 2: **Strain tuning results.** **A** Longitudinal resistivity $\rho_{aa}$ against *c*-axis field for sample #3, at a series of applied strains $\varepsilon_{aa}$. **H** = (0, 0, $H_c$), $T \approx 300$ mK. The measurement configuration is shown at top right. **B** Transverse response for the same sample. $V_{46}$ is the voltage between contacts 4 and 6, $V_{BG}$ the background, determined by fits to $V_{46}$ outside the region of phase formation, and $I_{35}$ is the current between contacts 3 and 5. (Contact numbering is shown in Fig. 1A.)

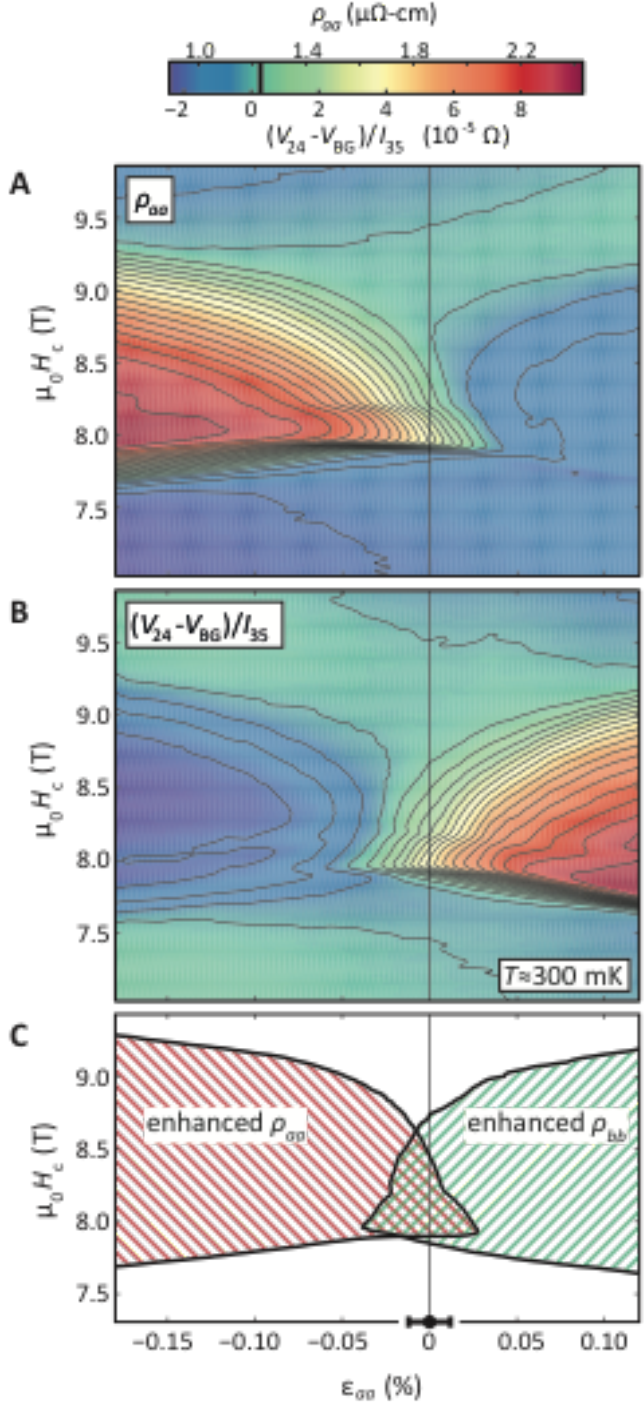

Figure 3: Strain tuning contour plots. **A** and **B**: Contour plots of the longitudinal and transverse response, respectively, of sample #3 as a function of *c*-axis magnetic field and sample strain. **H** = (0, 0, $H_c$), $T \approx 300$ mK. **C** Outlines of the regions of enhanced $\rho_{aa}$ and $\rho_{bb}$. These outlines are taken, for convenience, as the contour level indicated by the heavy black line in the scale bar. The error bar on the *x* axis indicates the error in our determination of zero strain. Although this phase diagram indicates a small asymmetry about zero strain that suggests spontaneous $C_4$ symmetry breaking, it should be treated with care. Within error the measurement is consistent with $C_4$ symmetry being respected at zero strain, and previous

measurements on entirely strain-free samples have shown $C_4$-symmetric transport in the absence of in-plane magnetic fields [11, 43].

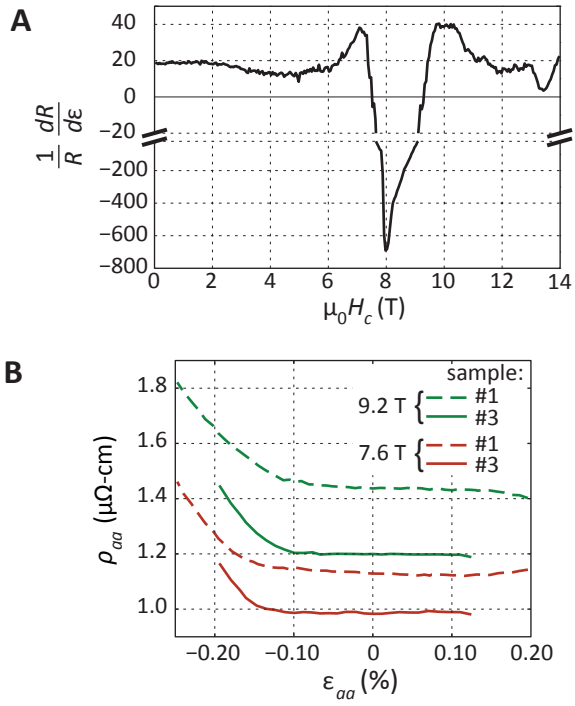

Figure 4: **Elastoresistance.** **A** Elastoresistance $(1/R)dR/d\varepsilon$ against magnetic field for sample #1. **H** = $(0, 0, H_c)$, $T \approx 300$ mK. **B** $\rho_{aa}$ against strain at 7.6 and 9.2 T. At these fields, $\rho_{aa}$ is nearly independent of strain over a wide range of strain.

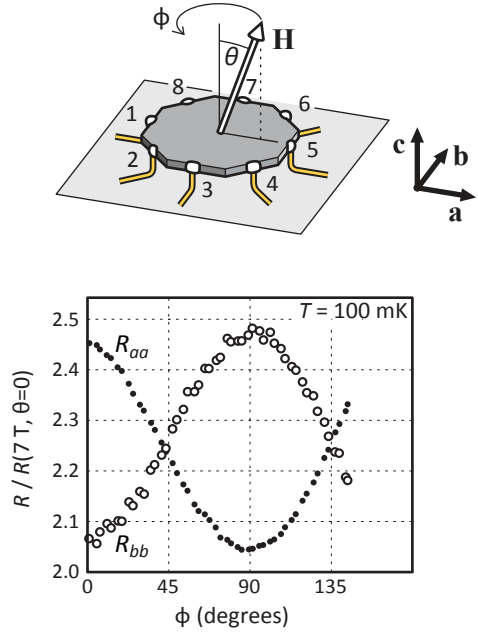

Figure 5: **Vector field sample configuration, and resistance against φ.** Top: sample and electrical contact configuration for the octagonal samples. Bottom: $R_{aa}$ and $R_{bb}$ against φ, with θ fixed at 6° and $\mu_0|\mathbf{H}|$ fixed at 7.83 T. $R_{aa} = (V_2 - V_4)/I_{15}$, and $R_{bb} = (V_4 - V_6)/I_{37}$.

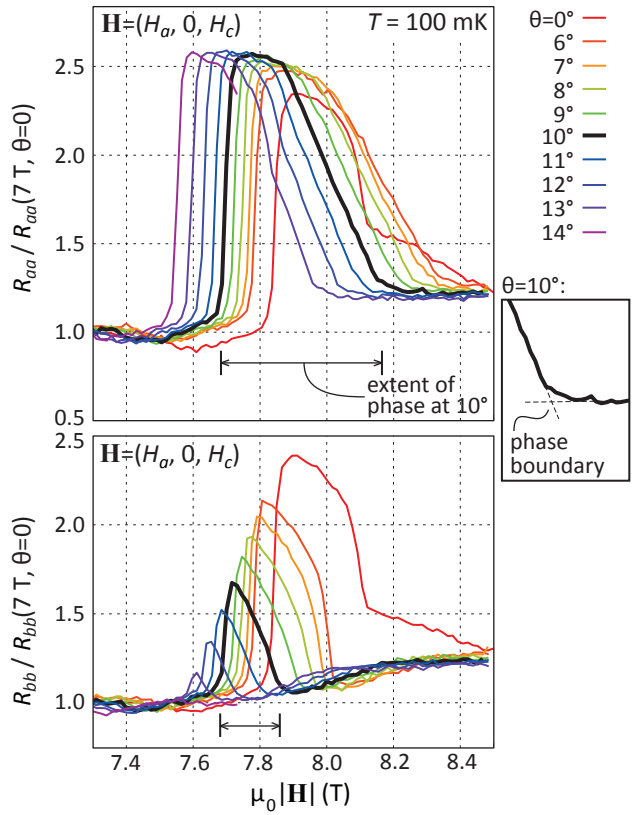

Figure 6: **Resistance against $\theta$ and |H|.** $R_{aa}$ and $R_{bb}$ against |H| at various field angles $\theta = \tan^{-1}(H_a/H_c)$. $H_b=0$, $T = 100$ mK. The phase boundaries are determined by the changes in slope, illustrated, as an example, for $\theta = 10°$.

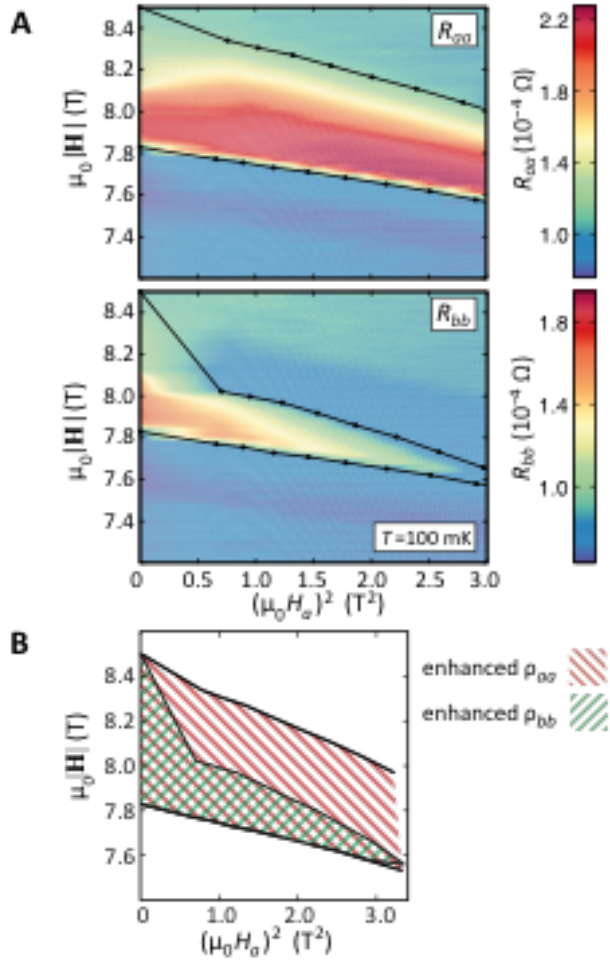

Figure 7: **Vector field results: contour plots. A** $R_{aa}$ and $R_{bb}$ against $|\mathbf{H}|$ and $H_a^2$, with $\mathbf{H} = (H_a, 0, H_c)$, at $T = 100$ mK. The black points are the transition points into enhanced-$\rho_{aa}$ and enhanced-$\rho_{bb}$ phases, determined as illustrated in Fig. 6. **B** Outlines of the regions of enhanced $\rho_{aa}$ and $\rho_{bb}$.

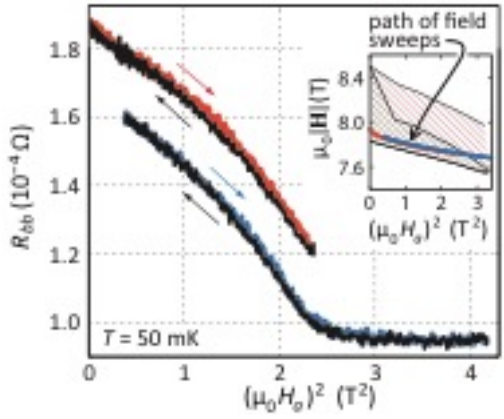

Figure 8: **Looking for hysteresis.** Two demonstrations of the absence of hysteresis, in which $R_{bb}$ was measured while increasing then decreasing $H_a$. The upper curves are offset by $2 \cdot 10^{-5}$ Ω for clarity. $I = 100$ μA, **H** = $(H_a, 0, H_c)$, $T = 50$ mK. As shown in the inset, $H_c$ was varied while $H_a$ was ramped to keep the sample within the enhanced-$\rho_{bb}$ phase up to $(\mu_0 H_a)^2 \approx 2.3$ T$^2$.

**Supplemental Material** for Brodsky *et al*, "Strain and Vector-Field Tuning of the Anomalous Phase in $Sr_3Ru_2O_7$."

In this Supplemental Material, we show data from strained sample #2, in Fig. S1. There are some technical differences from strained sample #3 in the transverse response: measurements were performed at a frequency were there was no substantial background that needed to be subtracted off, however the data were also noisier at this frequency. These comments aside, as with sample #3 (Fig. 4), there is a small but definite region of overlap of enhanced $\rho_{aa}$ and $\rho_{bb}$.

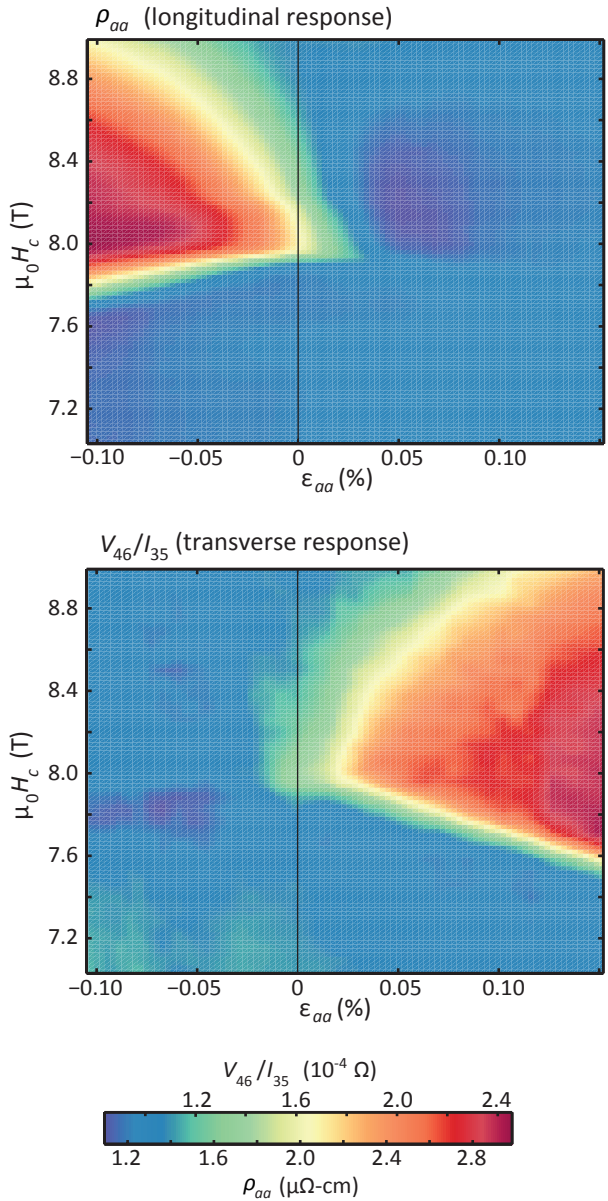

Figure S1: Longitudinal response and transverse response for strained sample #2.